\documentclass[preprint]{revtex4-1}
\usepackage{graphicx}
\usepackage{natbib}

\begin{document}
\newcommand{\micron}{$\rm{\mu m}$}
\newcommand{\Jcm}{$\rm{J/cm^2}$}
\newcommand{\mJcm}{$\rm{mJ/cm^2}$}
\newcommand{\microJ}{$\rm{\mu J}$}
\newcommand{\microSec}{$\rm{\mu s}$}

\title{Neural-network-assisted \textit{in situ} processing monitoring by speckle pattern observation}
\date{\today}
\author{Shuntaro Tani}
\email{stani@issp.u-tokyo.ac.jp}
\affiliation{Institute for Solid State Physics, The University of Tokyo}
\author{Yutsuki Aoyagi}
\affiliation{Institute for Solid State Physics, The University of Tokyo}
\author{Yohei Kobayashi}
\affiliation{Institute for Solid State Physics, The University of Tokyo}
\begin{abstract}
We propose a method to monitor the progress of laser processing using laser speckle patterns. Laser grooving and percussion drilling were performed using femtosecond laser pulses. The speckle patterns from a processing point were monitored with a high-speed camera and analyzed with a deep neural network. The deep neural network enabled us to extract multiple information from the speckle pattern without a need for analytical formulation. The trained neural network was able to predict the ablation depth with an uncertainty of 2 \micron, as well as the material under processing, which will be useful for composite material processing.
\end{abstract}
\maketitle
\section{Introduction}
Laser-based micro-structuring is a powerful tool for precision manufacturing and is considered a key element for future automized manufacturing. The main physical phenomenon behind laser-based micro-structuring is laser ablation: the process of surface material removal by rapid energy injection with laser pulses \cite{Chichkov:1996kf,Mueller:2014bc}. In particular, the use of ultrashort pulses enables material removal with negligible heat degradation. As such, much research has been devoted to properly control and understand this process. 

An important parameter in laser ablation is the rate of material removal, or the ablation rate. The ablation rate depends on various factors, such as material properties, surface morphology, pulse energy, and the laser repetition rate \cite{Nolte:1997jk,Hashida:2002hy,Ancona:2008ec,Tani:2018cg, Mustafa:2020hl}. For precision manufacturing and automized process optimization, real-time monitoring of this processing rate is of great importance. Specifically, the monitoring of the cumulative ablation rate, or the processed depth or volume, is important for hole drilling or groove processing.

Various methods have been developed to monitor the processing status in a non-contact way, including acoustic and plasma detection \cite{Yeack:1998cu,Sun:1999hk,Stournaras:2010fq,Stournaras:2010gd}. Specifically, laser interferometry is a powerful technique to quantify the progress of laser processing. In laser interferometry, a monitoring beam is divided into two paths: one for the actual monitoring, and another as a reference. The monitoring laser beam is projected onto a processing point, and the reflection of this light is directed back into an interferometer. The interference signal between this beam and the reference is measured, where the signal strength depends on the relationship between the optical path lengths of the two beams.
By properly tracking changes in this signal during processing, it is possible to extract  real-time processed depth information. Intensive studies have been performed to increase the precision and acquisition rate of laser interferometry setups, as well as to make the overall system convenient to install. Currently, there are two major implementations, differentiated by their choice of reference beam path: cross-interferometry and self-mixing.
In cross-interferometry, a separate reference arm is placed in the interferometer, which enables the determination of absolute distance in exchange for a comparatively large installation footprint \cite{Webster:2011ji,Webster:2014bu,Blecher:2014bx,Ji:2015cy,Stadter:2020dn}. In self-mixing-interferometry, the intact surface outside the ablation region is used for the reference path \cite{Mezzapesa:2012jx,Mezzapesa:2012ev,Demir:2015ik}.  Although the self-interferometry method does not require a reference arm, the counting of net oscillation is necessary to quantify the total ablated depth, limiting its uses to situations with gradual depth changes, such as percussion drilling.

As an alternative method to optically monitor the laser processing status, we focus on utilizing the spatial coherence of a probe light source instead of only the temporal coherence as in the aforementioned interferometry techniques. When spatially coherent light is scattered by a rough surface and projected onto a screen, the scattered light on the screen exhibits a highly irregular intensity profile, known as a laser speckle pattern. The speckle pattern arises from interference of wavefronts scattered at different positions of the surface. As such, the speckle pattern carries a trove of information on the material surface condition. It has been used for various measurement and diagnostic applications, such as in surface roughness measurements \cite{Leger:1975ej}, flow measurements\cite{DUDDERAR:1977dl, Dunn:2016dj}, and strain measurements\cite{Yamaguchi:1981bd}. Furthermore, techniques for three-dimensional reconstruction from the speckle pattern have been developed, realizing sub-millimeter precision \cite{Yamamoto:1994je, Dekiff:2010er}. As speckle-pattern analysis only requires illumination and an observation camera, it should be straightforward to implement into laser processing setups. Moreover, the two-dimensional nature of the speckle pattern observation should provide additional information compared to conventional one-point interferometric measurements. The major challenge here then becomes extracting meaningful physical characteristics from the complex speckle pattern.

\begin{figure}
  \includegraphics[width = 8cm]{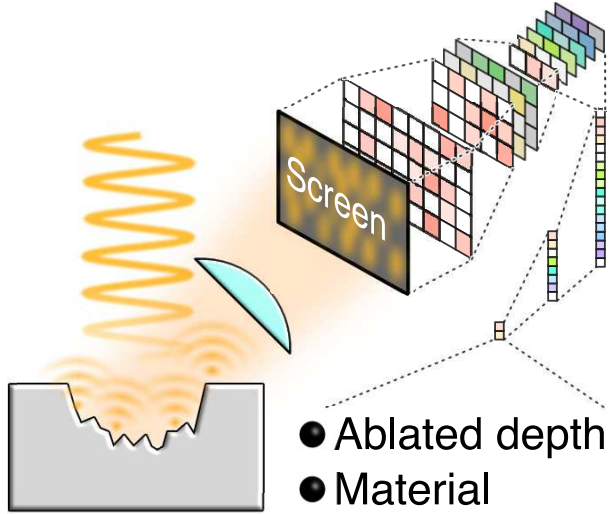}
  \caption{Process monitoring using laser speckle with a deep neural network.}
\end{figure}
In this study, we demonstrated the successful extraction of multiple information from three sequential frames of speckle patterns measured \textit{in situ} during laser processing. Figure 1 shows the concept of our method. We illuminate the processed surface with a coherent light source, and observe the reflected speckle pattern with a high-speed camera. Traditionally, the quantification of speckle patterns required a formulation for application- and measurement- specific analysis. Here, we utilize neural networks to practically overcome this difficulty. Deep learning are emerging technology to ties multi-dimensional vectors to other multi-dimensional vectors with the neural networks using a number of datasets \cite{Krizhevsky:2012wl,LeCun:2015dt,Tian:2018gs,Turpin:2018bo,Muskens:2019gi,Ozcan:2019fa,Kalichman:2019hu} . We use a number of experimental datasets to train a neural network to approximate the complex nonlinear functions required to extract meaningful physical values from a speckle image. We show that a deep neural network can use speckle patterns to extract multiple information regarding the processing status during laser processing, such as ablated material type, depth, and volume. The simple setup should be a great candidate for next-generation process monitoring technology. 
\section{Experimental setup}
\begin{figure}
  \includegraphics[width = 8cm]{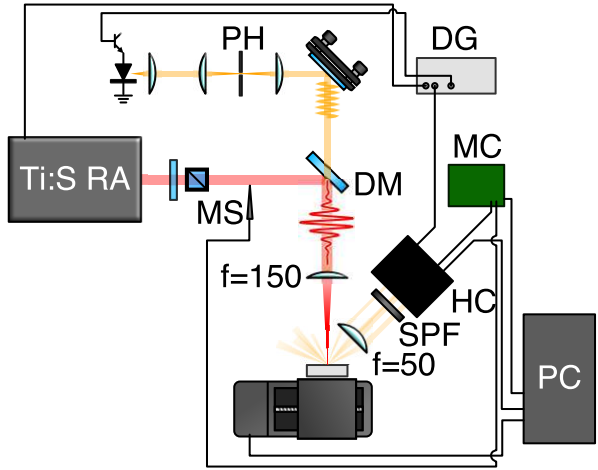}
  \caption{Experimental setup. The abbreviations are as follows: PH: pinhole, DG: delay generator, MS: mechanical shutter, DM: dichroic mirror, MC: micro-computer, HC: high-speed camera, and SPF: short pass filter.}
\end{figure}
The experimental setup is shown in Fig. 2. We used plates of aluminum, copper, and nickel for the sample. The roughness Ra of the plates before the processing was measured to be typically around 1 \micron. The sample was placed on a two-axis stage, which moved at a constant velocity during groove processing. Laser pulses for rapid surface engraving were provided by a Ti: Sapphire regenerative amplifier with a wavelength of 800 nm, pulse duration of 35 fs, and a repetition rate of 1 kHz. The pulse energy of the laser pulses was adjusted by a motor-driven half-wave plate and a polarizing beam splitter.
The laser pulses were then focused onto the sample surface through a lens with a focal length of 150 mm; the spot size on the target was approximately 15 \micron. A mechanical shutter was used to switch on and off the laser irradiation. A laser beam for monitoring was provided from a laser diode with a wavelength of 633 nm. The monitoring beam was spatially filtered through a 50-\micron pinhole, and then coaxially merged with the femtosecond laser pulses through a dichroic mirror.
The spot size on the target was 20 \micron, which was 1.5 times larger than that of the processing beam. Laser speckle patterns from the target surface were recorded with a high-speed camera through a collector lens with a focal length of 50 mm. The lens was placed between the target and the high-speed camera to match the Fourier plane of the lens with the detector plane. A laser line filter at the wavelength of 632.8 nm was inserted in between the lens and the camera.
The high-speed camera was synchronized with a 1-kHz TTL signal from the regenerative amplifier. The high-speed camera was also synchronized with the mechanical shutter, first nine frames without laser irradiation due to a delayed response of the mechanical shutter. The exposure time was set to 500 \microSec; this exposure was also synchronized with the monitoring laser diode. The acquired image frames were sent to the computer in real time through CoaXPress cables. The pulse energies were varied from 10 to 60 \microJ, and the scan speeds were varied from 1.0 to 3.5 mm/s.
\begin{figure}
  \includegraphics[width = 16cm]{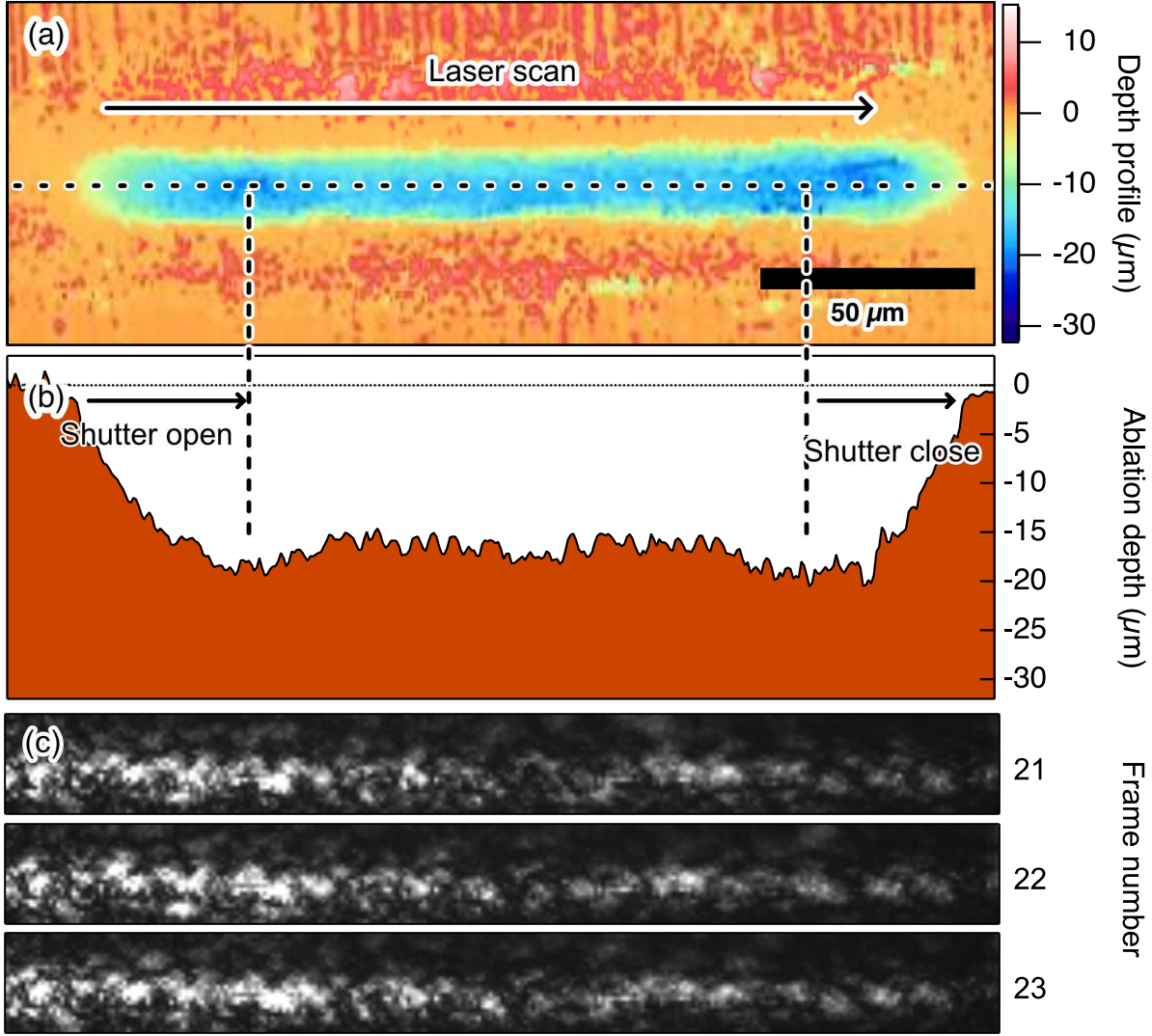}
  \caption{(a) Three-dimensional depth profile of a laser processed groove. (b) Cross-sectional of the depth profile. (c) Typical laser speckle images.}
\end{figure}
The depth profiles of the grooves were measured using a confocal 3D microscope with a 50x objective lens. A typical depth profile is shown in Fig. 3 (a), where the laser pulses were swept along the horizontal axis. The cross-section of the depth profile along the laser trace is shown in Fig. 3(b). When illuminated by the monitoring diode, the macroscopic groove shape and the microscopic groove surface roughness combine to form the total laser speckle pattern. Figure 3(c) shows typical sequential images of the observed speckle patterns. As the processing proceeds, the changes in the surface structure and the monitoring position induce changes in the laser speckle patterns. The diffraction angles for the left and right edge of the images are 30 and 60 degrees, respectively.
\section {Data preparation and training}
The resolution of the acquired images was 4080 x 480 pixels, which was reduced down to 255 x 25 pixels by box averaging and cropping. The brightness and contrast of each image were normalized between 0 and 1. Due to this normalization, only spatial patterns of the images were taken into account in the following analysis. The neural network consisted of two convolutional layers, five residual convolutional layers, and a max-pooling layer followed by two convolutional layers. The last two-dimensional layer was then flattened and connected to two fully connected layers. The rectified linear unit was used for the activation function. Three sequential image frames were used as input data, and the measurement value and material vectors were used as label data. Here the material vector was represented by a one-hot vector of material species. The loss function was composed from the root mean square of the ablated depth and the cross-entropy of the material vector. The neural network was trained with 13,000 datasets which included all three materials. The batch size was 256, and the Adam algorithm was used for optimization. 
\section {Results}
\begin{figure}
  \includegraphics[width = 16cm]{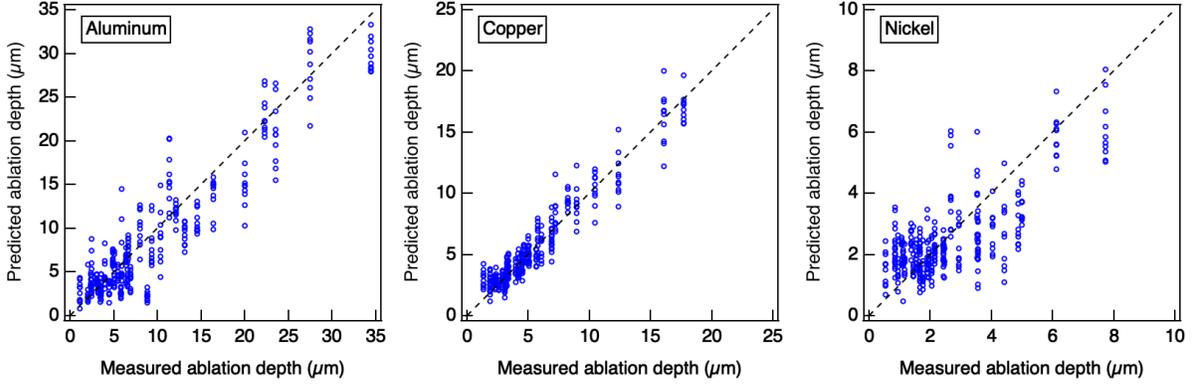}
  \caption{Predicted ablation depth vs. measured ablation depth for aluminum, copper, and nickel.}
\end{figure}
Upon three-sequential image input, the trained neural network predicts the ablation depth and the target material. Figure 4 shows the validation results of the trained neural network. Validation data were obtained separately from the training data. The 32-34 combination of pulse energies and scan speed for the three materials resulted in 98 datasets. Each dataset consisted of 10 three-sequential images. The net 980 datasets were used for the validation. The estimated ablation depth by the neural network was plotted as a function of the experimentally measured ablation depth. Good correspondence was found between estimated and measured values, with less than 2 \micron uncertainty and without material dependence. Most of the dataset showed less than 1 \micron uncertainty below an ablation depth of 5 \micron. We note that the prediction took 0.09 ms using a standard commercially available GPU; this figure shows that with proper configuration of the camera data acquisition board and GPU memory, real-time feedback of the laser processing should be possible.
\begin{figure}
  \includegraphics[width = 8cm]{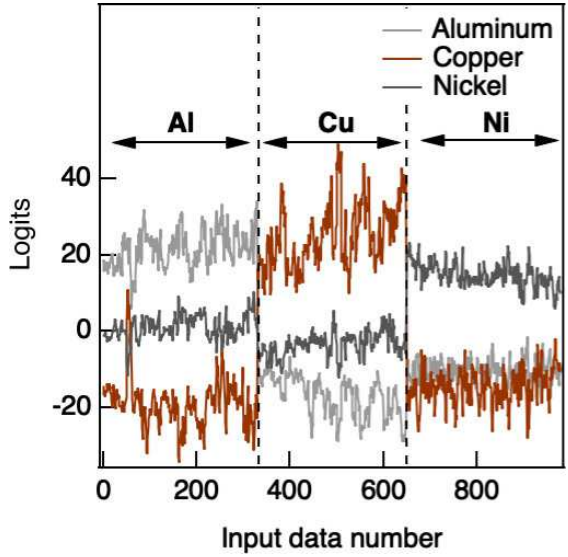}
  \caption{Logits magnitude for various input frames for aluminum, copper, and nickel.}
\end{figure}
In addition, the speckle patterns can simultaneously predict the material type being processed. Figure 5 shows the predicted logits of materials, where the logit is defined as log p – log(1-p), where the p is the probability of being the material. A high logits value output indicates a high probability of the input being a certain material, and vice-versa for a low value. The horizontal axis indicates the indices of the validation datasets, where the correct material type for a particular index range is indicated by the overlaid horizontal arrows. As shown in the figure, the predicted material vector shows good agreement with the material under processing. Almost 10 dB separation between the most probable material and the others shows that we can accurately predict the material under processing. The good material separation for various ablation depths suggests the existence of material-specific surface morphology for each groove, that this information is imprinted in the speckle pattern, and that the deep neural network can successfully extract this information.
\begin{figure}
  \includegraphics[width = 16cm]{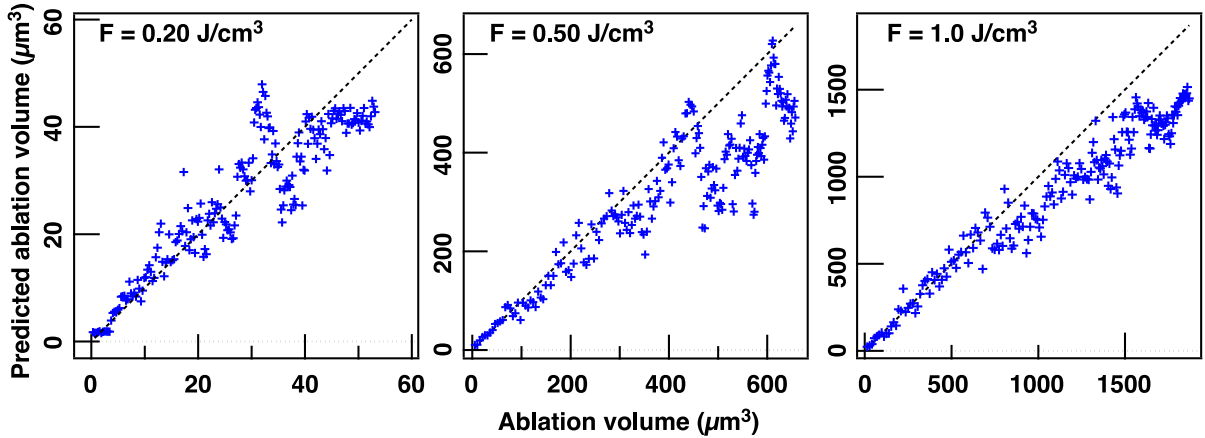}
  \caption{Deep-neural-network based prediction of ablation volume from laser speckle pattern during percussion drilling.}
\end{figure}
Our method can also be applied for hole drilling. Figure 6 shows the predicted ablation volume as a function of measured ablation volume in percussion drilling. In this experiment, we used polished plates of silicon as a target material. The same experimental setup was used without moving the stage during laser irradiation. Holes were drilled with various combinations of pulse energy and the number of pulses. The number of pulses ranged from 50 to 300 with a step size of 50, which was controlled by the mechanical shutter. The pulse energies were varied from 1 to 10 \microJ. Ablated volumes of the drilled holes were measured using the 3D confocal microscope. We prepared the label data by interpolating the ablated volume as a function of pulse energy and the irradiated number of pulses. The image frames for the training and validation were prepared independently. Almost 100,000 datasets were used for the training and 1,000 datasets were used for the validation. The good correspondence between the predicted value and the actual value suggests that our method is also applicable for drilling.
\section {Conclusion}
We developed a method to extract the progress of laser processing by \textit{in situ} speckle pattern analysis using a neural network. The neural network can predict the ablation depth of a groove with an accuracy of 2 \micron. Our method is easy to install compared to traditional interferometric monitoring techniques. We also show that as the speckle information utilizes the full surface information, it can predict other properties of the processed material as well, for example, the processed material type. This would be especially useful for composite material processing. Moreover, the prediction time of less than 0.1 ms is promising for real-time feedback for complex processing procedures. Altogether, the simplicity, versatility, and overall accuracy of the method should make it a strong candidate for future integrated monitoring systems. 
\section* {Funding}
A part of this word is supported by a TACMI project (NEDO).

\section* {Acknowledgements}
The author would like to thank Dr. Haruyuki Sakurai for constructive criticism of the manuscript.
\section* {References}

\end{document}